\documentclass[pra,twocolumn,aps,10pt,superscriptaddress]{revtex4-2}
\usepackage{hyperref,xcolor, comment, graphicx, physics}
\usepackage{amssymb,amsmath,bm,units}
\usepackage{upgreek,bm,booktabs}

\begin{document}
\title{Comparing sliding-mode, bang–bang
and linear-quadratic-Gaussian \\
for steering an atomic clock}
\author{Ashkan Bayat\href{https://orcid.org/0000-0001-8104-5937}{\includegraphics[scale=0.05]{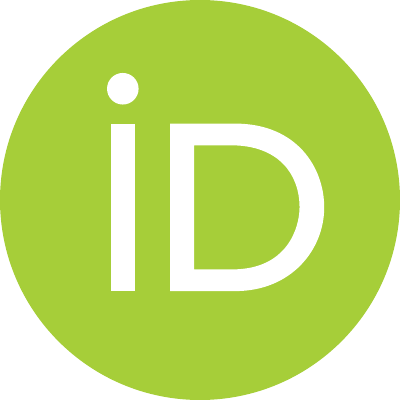}}}
\email{abayat2@ualberta.ca}
\affiliation{Institute for Quantum Science and Technology, University of Calgary, Calgary, Alberta, Canada T2N 1N4}
\affiliation{University of Alberta, Edmonton, Alberta, Canada T6G 2R3}
\author{Barry C. Sanders\href{https://orcid.org/0000-0002-8326-8912}{\includegraphics[scale=0.05]{figures/orcidid.pdf}}}
\email{sandersb@ucalgary.ca}
\affiliation{Institute for Quantum Science and Technology, University of Calgary, Calgary, Alberta, Canada T2N 1N4}
\begin{abstract}
Accurate timekeeping relies on feedback that continually steers a local clock toward a higher-grade reference. 
We evaluate first-order sliding-mode control (SMC) for steering an atomic clock and benchmark it against two standards: linear–quadratic–Gaussian (LQG) control and the bang-bang~(BB). 
All three are tested in a common numerical framework using the standard two-state clock model driven by white and random-walk-frequency noise. 
To ensure the conclusions are not tied to a single noise realization and a single time period, we repeat the accuracy analysis over 100 independent random seeds for four different time periods, reusing the same seed across controllers within each trial.
The time periods considered are one week, one month, one year, and ten years to cover short-, mid-, and long-term analyses of accuracy. 
Our results show that SMC remains competitive with LQG across the tested timescales and reference-clock qualities. 
Both SMC and LQG substantially outperform BB over the same time periods.
Over the full averaging-time range studied, SMC’s stability is almost identical to LQG’s, whereas BB shows the characteristic short-term instability. 
Together, our results indicate that SMC is a promising clock-steering policy that can remain close to LQG in accuracy while avoiding the short-term instability seen in BB.
\end{abstract}
\date{\today}
\maketitle
\section{Introduction}
Precise time and frequency references are used for navigation, telecommunications, radio astronomy, and fundamental-physics experiments~\cite{KM23,KLS+21,HBO+18,PSS+21, GCGL24}. 
Virtually every modern system therefore embeds a steering loop that nudges a local clock toward an external or ensemble reference so that its time error stays within nanoseconds over months to years~\cite{ZWW+24}. 
Early work introduced simple on-off---``BB” policy---corrections to keep the Global Positioning System (GPS)
aligned with the local realization of Coordinated Universal Time~(UTC) by the United States Naval Observatory, abbreviated as UTC(USNO)~\cite{Cha94}.

Over the past two decades, more sophisticated designs have been adopted~\cite{YPL17}. 
The linear-quadratic-Gaussian (LQG) scheme, first demonstrated on hydrogen masers at USNO in 1995~\cite{KL95} and later analyzed in depth by Farina, Galleani, Tavella \&\ Bittanti in 2010 (FGTB10)~\cite{FGTB10}, delivers optimum linear–quadratic feedback law by weighting time, frequency, and control-effort cost in a Riccati-based feedback law. 
At the opposite end of the complexity spectrum is the GPS BB. 
The BB needs only a sign comparison, although its aggressive switching injects extra noise that appears as a characteristic bump in Allan deviation plots as shown by FGTB10.

Sliding mode control~(SMC) is a natural middle ground:
it maintains the simplicity of BB while inheriting much of LQG’s robustness. 
Yet, published applications of SMC to atomic clocks are confined to short, acquisition-only stages. 
For example, consider the two-stage fast-synchronization algorithms recently proposed for
oscillators disciplined by the global navigation satellite system (GNSS).
This stage-one algorithm disables SMC once the time-offset is below a certain threshold~\cite{KJK25}.
Here we analyze a continuous SMC loop as the primary long-term steering policy and compare it side-by-side with LQG and BB under identical stochastic conditions.

We use the standard two-state clock model driven by white-frequency noise~(WFN) and random-walk-frequency-noise~(RWFN).
Then we simulate
\begin{itemize}
\item an LQG-steered clock,
\item a GPS-style BB-steered clock,
\item a continuously active first-order SMC-steered clock, and
\item an unsteered one.
\end{itemize}
Next, we perform two complementary simulations: (i)~time-domain runs that provide time-offset records for accuracy evaluation and (ii)~a million-day~($\approx 10^{11}\, \text{s}$) run that supplies a sufficiently long data set for clean overlapping Allan-deviation (oADEV) estimates.
For accuracy, we repeat the time-domain simulation over 100 different random seeds and evaluate the accuracy over four time periods:
\begin{enumerate}
\item one week $\approx 6.05\times10^{5}\, \text{s}$,
\item one month $\approx 2.6\times 10^{6}\, \text{s}$, \item one year $\approx 3.15\times10^{7}\, \text{s}$, and
\item ten years $\approx 3.15\times10^{8}\, \text{s}$.
\end{enumerate}
We reuse the same seed number across all the controllers for each evaluation for a fair comparison.

To evaluate stability, we use only one seed and use the oADEV as the performance metric. 
We evaluate the oADEV over 
\begin{equation}
\label{eq:Allandeviation}
10^5~\text{s} \leq \tau \leq 10^8~\text{s}.
\end{equation}
Our results show that the SMC, tuned with one pair of parameters
\begin{equation}
\label{eq}
\lambda = 6\times 10^{-6}~\text{s}^{-1},\;
K = 1.1\times 10^{-19}~\text{s}^{-1},
\end{equation}
remains competitive with LQG across the tested timescales and reference-clock qualities without the short-term instability seen in BB.

Our paper is organized as follows. 
We present essential background in~\S\ref{sec:back}.
We introduce the two‑state clock model in~\S\ref{subsec: tscm}.
Next, we outline the model for controlling a clock in \S\ref{subsec:cm}.
Subsequently,
\S\ref{subsec:cs} reviews the three steering policies:
LQG, BB, and first‑order SMC.
Then, in~\S\ref{sec:approach},
we describe our approach on how we compare the three steering policies for an atomic clock,
and present our numerical results in \S\ref{sec:res}.
We discuss our results in \S\ref{sec:dis} and conclude  in \S\ref{sec:con}.
For convenience,
we summarise our abbreviations in Table~\ref{tab:abbrev}.
\begin{table}
\centering
\begin{tabular}{|l|l|}
\hline
ADEV&Allen deviation\\
BB & bang-bang \\
DARE&discrete-time algebra\"{i}c Riccati equation\\
FGTB10&Farina, Galleani, Tavella \&\ Bittanti 2010\\
FRC&free-running clock\\
GNSS&global navigation satellite system\\
GPS&global positioning system\\
LQG&linear-quadratic-Gaussian\\
oADEV&overlapping ADEV\\
RNG&random number generator\\
RWFN&random-walk frequency noise\\
SMC&sliding mode control\\
USNO&United States Naval Observatory\\
UTC&coordinated universal time\\
WFN&white frequency noise\\
\hline
\end{tabular}
\caption{Summary of abbreviations in alphabetical order.}
\label{tab:abbrev}
\end{table}
\section{Background}
\label{sec:back}
In this section, we establish the theoretical underpinnings for our comparative study. 
Below,
\S\ref{subsec: tscm} sets out the standard two‑state clock model driven by independent WFN and RWFN and states the discrete‑time form used in all simulations. 
We then explain in \S\ref{subsec:cm} the control model used throughout this paper.
Next, \S\ref{subsec:cs} summarizes the two main steering strategies used to steer a clock: The LQG and the GPS BB.
Finally, we describe the first‑order sliding‑mode controller, which is essential for the subsequent sections, where we apply this idea to clock steering.
\subsection{Two-state clock model}
\label{subsec: tscm}
In this subsection we first define the clock states (time offset and fractional-frequency error).
Then, we outline the continuous-time stochastic model we adopt.
Finally, we provide the discrete form of the stochastic model that serves as the plant for all controllers later. 

Now we explain the two-state model.
We model a free-running clock~(FRC)
with two state variables
\begin{equation}
\label{eq:X}
    \bm{X}(t) = 
    \begin{pmatrix}
    X_1(t)\\
    X_2(t)
    \end{pmatrix}
\end{equation}
such that~$X_1(t)$
denotes the time offset with respect to a reference clock and
\begin{equation}
\label{eq:X2}
X_2(t):=\dot{X}_1(t)
\end{equation}
is the fractional frequency error~\cite{PT08}. 
Furthermore,
the standard stochastic model is a linear system driven by two independent Wiener measures
\begin{equation}
\label{eq:W}
    \text{d}\bm{W}(t) = 
    \begin{pmatrix}
        \text{d}W_1(t)\\
        \text{d}W_2(t)
    \end{pmatrix},
\end{equation}
which captures the dominant noise types used in timekeeping, namely, WFN and RWFN. 

Next we express and explain the stochastic dynamical equation for the FRC.
Using the free-clock variables~(\ref{eq:X}) 
and stochastic terms~(\ref{eq:W}),
we write
\begin{align}
\label{eq:cd}
    \text{d}{\bm{X}}(t) = F\,\bm{X}(t)\,\text{d}t + \bm{B}\,\text{d}\bm{W}(t)
\end{align}
for the continuous-time dynamics.
Here
\begin{equation}
\label{eq:matrixd}
    F := 
    \begin{pmatrix}
    0 & 1\\
    0 & 0
    \end{pmatrix},\,
    \bm{B} := \text{diag}(\sigma_1, \sigma_2),
\end{equation} 
which encode the deterministic kinematics and the scaling of the time and frequency noise, respectively. Intuitively, $\sigma_1$ quantifies the diffusion of time offset,
whereas ~$\sigma_2$ quantifies the diffusion of the fractional frequency-offset.
For simulation, we discretize time with sampling interval~$\tau_0$ so that elapsed time is 
\begin{equation}
\label{eq:kdeltat}
k\,\tau_0,\,
k\in\{0,\dots,N-1\}.
\end{equation}
Then, 
\begin{align}
\label{eq:dd}
    \bm{X}_{k+1} = A\,\bm{X}_k+\bm{B}\,\text{d}\bm{W}_k,\; 
\end{align}
with
\begin{equation}
\label{eq:Xk}
     A := 
    \begin{pmatrix}
    1 & \tau_0\\
    0 & 1
    \end{pmatrix},\;
    \bm{X}_k := 
    \begin{pmatrix}
    X_{1,\,k}\\
    X_{2,\, k}
    \end{pmatrix}
\end{equation}
replaces Eq.~(\ref{eq:cd})
as the discrete-time version of stochastic dynamics for the FRC.

\subsection{Control model}
\label{subsec:cm}
We proceed to outline the model for controlling a clock.
We describe control in the context of standard control theory, with the agents being the plant, the evaluator, and the controller~\cite{AM08}.
We then describe the dynamics of a steered clock.
Finally, we explain how the controller steers the clock under three strategies we study~(LQG, BB, and SMC).

In control theory,
the entity being controlled is the `plant',
which in our case is the FRC whose variables are represented by vector $\bm{X}$~(\ref{eq:X}).
The plant has two knobs for control and one sensor for detecting and transmitting information. 
One of these two knobs changes the clock’s fractional frequency~$X_2(t)$~(\ref{eq:X}), while the other knob applies a steady frequency drift~$\dot{X}_2$~\cite{PT08}. 
The sensor is a digital cycle counter that counts the oscillations and emits a signal being one pulse emitted each day,
which is abbreviated to
1~PPD.
Each day the signal pulse occurs after
\begin{equation}
\label{eq:cyclespersecond}
86400 \times 9,192,631,770
\end{equation}
cycles for a Cs clock.

Our evaluator in the control-theory model possesses a reference clock.
The evaluator receives 1~PPD from the plant and 1~PPD from the reference clock.
The evaluator then uses a time-interval counter, which starts
counting on receiving one pulse and stops on receiving the other pulse. 
The elapsed time that the time interval counter reports is the time-offset~$X_1$,
defined to be the interval
between the pulse arriving from the plant and the pulse arriving from the reference clock.
From these time-offset data, the evaluator then estimates the fractional-frequency offset~$X_2$ by subtracting consecutive~$X_1$ measurements and dividing by the update interval~$\tau_0$, which is one day in our case.
Finally, the evaluator sends~$X_1$ and~$X_2$ to the controller.

The controller receives information, namely~$X_1$ and~$X_2$ from the evaluator. 
These two offsets are inputs to the control policy that is used to enhance the clock performance.
From these inputs, the controller executes a command~$U_k$ according to a chosen policy,
e.g.,
\begin{equation}
\label{eq:policies}
\text{LQG}, \text{BB}, \text{SMC}.
\end{equation}
This command is either change the frequency of the clock by a certain amount if the policy is~LQG
or change the frequency drift (rate-of-change) if the policy is either BB or SMC.
We associate each of these types of command with either of the plant’s two control knobs:
for LQG, the controller uses the~$X_2$ knob~(\ref{eq:X2}), whereas for BB and SMC, the controller uses the~$\dot{X_2}$ knob.

We introduce the discrete-time dynamics of a steered clock.
Let~$\bm{C}$ be an input vector that determines which knob is used by the controller.
The discrete-time dynamics~(\ref{eq:dd}) is modified to
\begin{equation}
\label{eq:sdd}
\bm{X}_{k+1} = A\,\bm{X}_k+\bm{C}\,U_k+\bm{B}\,\text{d}\bm{W}_k
\end{equation}
to represent the dynamics of a steered clock.

The next step is to connect $\bm{C}$ to the knob the controller uses.
Choosing
\begin{equation}
\label{eq:C_LQG}
\bm{C}^{X_2} :=
\begin{pmatrix}
    \tau_0\\
    1
\end{pmatrix}
\end{equation}
selects the $X_2$ knob, meaning~$U_k$ acts directly on~$X_2$ and integrates into~$X_1$.
The LQG policy employs the input vector~$\bm{C}$~(\ref{eq:C_LQG}).
Choosing
\begin{equation}
\label{eq:C_BBSMC}
\bm{C}^{\dot{X}_2} := 
\begin{pmatrix}
    \tau_0^2/2\\
    \tau_0
\end{pmatrix}
\end{equation}
selects the $\dot X_2$ knob, meaning~$U_k$ accumulates as a rate in $X_2$ and, in turn, integrates into~$X_1$ across the interval.
The BB and SMC policies employ the input vector given by Eq.~(\ref{eq:C_BBSMC}).

\subsection{Control strategies}
\label{subsec:cs}
Now we elaborate on the three control strategies used in our study. 
We begin with LQG, explaining why it is called quadratic in the first place. Then we derive the feedback law used in LQG and show why LQG is called linear.
Next, we discuss BB and SMC. Finally, we explain the two metrics used for evaluating and comparing the steering policies' performance.

First, we focus on the quadratic part of LQG.
For LQG we seek a control law that minimizes a quadratic cost function~\cite{FGTB10}. The cost function being quadratic establishes the ``Q" in LQG.
The cost function consists of two weights.
The first weight is a $2\times2$ symmetric state-weighting matrix~$W_{Q}$ and the second weight is a scalar~$W_R$.

As a first step, we focus on how~$W_\text{Q}$ and~$W_R$ contribute to the cost function.
The contribution of~$W_Q$ to the cost function is the quadratic form
\begin{equation}
\label{eq:XW_Q}
\bm{X}_k^{T}W_Q\bm{X}_k,
\end{equation}
with the cross terms in~$W_Q$ usually set to zero.
The diagonal elements of~$W_Q$ set the penalty weight on~$X_1$ and~$X_2$, respectively. 
The contribution of~$W_R$ to the cost function is in the form
\begin{equation}
\label{eq:WR}
    U_kW_RU_k,
\end{equation}
which penalizes the control command.
Given the weights~$W_Q$ and~$W_R$, the quadratic cost function is~\cite{FGTB10}
\begin{equation}
    \label{eq:cf}
    J_{\text{opt}} = \mathbb{E}\left[\sum_{k = 1}^{\infty}\left(\bm{X}_k^{T}W_Q\bm{X}_k+U_kW_RU_k\right)\right],
\end{equation}
which drives~$X_{1,2}$ to small values via~$W_Q$.
The second term discourages large control commands via~$W_R$.
Tuning $\left(W_Q, W_R\right)$ sets the trade-off between accuracy and stability.

We now derive the feedback law for minimizing $J_{\text{opt}}$~(\ref{eq:cf}), which also clarifies why LQG is called linear.
First, we solve the discrete-time algebra\"{i}c Riccati equation
(DARE)~\cite{KJK25}
\begin{align}
\label{eq:Riccati}
    P =& A^TPA
    -\left(A^TP\bm{C}^{X_2}\right)
    \left(W_R+\left(\bm{C}^{X_2}\right)^TP\bm{C}^{X_2}\right)^{-1}
    \nonumber\\&
\times
\left(\left(\bm{C}^{X_2}\right)^TPA\right)
+W_Q,
\end{align}
for~$P$ the solution of the DARE.
Solving the DARE yields the gain vector
\begin{equation}
\label{eq:LQGgain}
    \bm{K}^\wp = \left(W_R+\bm{C}^TP\bm{C}\right)^{-1}\left(\bm{C}^TPA\right)
\end{equation}
for~$\wp$
a label we use for any of the three policies~(\ref{eq:policies});
in this case
$\wp\gets\text{LQG}$.
For a linear plant model (\ref{eq:sdd}), the optimal regulator that minimizes the quadratic cost function~(\ref{eq:cf}) is a linear state-feedback law
\begin{equation}
\label{eq:LQGU}
    U^{\text{LQG}}_k = -\bm{K}^{\text{LQG}}\, \cdot \bm{X}_k,
\end{equation}
which establishes the ``L" in LQG. 

Now we summarize the GPS-heritage policy, namely, BB.
In this policy, the GPS master control station steers satellite clocks using the BB control strategy that looks only at the sign of a sliding surface 
\begin{equation}
\label{eq:BBss}
S^{\text{BB}}_k := X_{1,\, k}+\frac{\left(X_{2,\,k}\right)^2\text{sign}\left(X_{2,\,k}\right)}{2K_{\text{BB}}},
\end{equation}
which is a surface that represents confinement of state vectors~(\ref{eq:X}).
The sliding surface combines~$X_{1k}$ and~$X_{2k}$ into a single measure that the controller tries to drive to zero.
The feedback law is 
\begin{equation}
\label{eq:BBu}
U_k = - K_{\text{BB}}\,\text{sign}\left(S^{\text{BB}}_k\right),
\end{equation}
where the constant $K_{\text{BB}}$ sets the magnitude of the control step. 
The policy is attractive because the entire decision is based on one sign comparison.

We next introduce first-order SMC, which is used as a clock-steering strategy in this paper.
Like the BB, SMC uses a sliding surface and a discontinuous feedback law~\cite{utk77}.
In fact, the BB can be viewed as a simple sliding-mode scheme with a different sliding surface and a purely on/off control term.
The classical first-order SMC design introduces a linear sliding surface
\begin{equation}
\label{eq:SMCs}
S^{\text{SMC}}_k := X_{2,\,k}+\lambda X_{1,\,k},
\end{equation}
and the associated feedback law
\begin{equation}
\label{eq:SMCu}
U_k = -\lambda X_{2,\,k} - K_{\text{SMC}}\,\text{sign}\left(S^{\text{SMC}}_k\right)
\end{equation}
guarantees finite-time convergence of~$S^{\text{SMC}}_k$ to zero.
Here, $K_{\text{SMC}}$ sets the size of the control step and $\lambda > 0$ quantifies
the~$X_1(t)$ decay rate once the trajectory is sliding on~$S_{\text{SMC}}$.

To compare the three steering policies, we use two performance metrics: accuracy and stability~\cite{Lom02}.
Accuracy is how closely the clock’s time stays aligned with the reference and is quantified by the standard deviation of the time-offset time-series data~\cite{AGM74}
\begin{equation}
\label{eq:sigmax1}
\sigma_{X_1} := \sqrt{\expval{X_1^2} - \expval{X_1}^2},
\end{equation}
where~$\langle\rangle$ denotes mean.
Stability is how the clock’s frequency fluctuates over time and is quantified using the overlapping Allan deviation (oADEV)~\cite{RH08}
\begin{align}
\label{eq:sigmaytau}
\sigma^2_y(n\tau_0) \approx &\frac1{2n^2\tau_0^2(N-2n)}
\nonumber\\&
\times\sum_{k = 0}^{N - 2n - 1}(X_{1,\,k+2n}
- 2X_{1,\, k+n}+ X_{1,\,k})^2,
\end{align}
for~$N$ the number of~$X_{1, k}$ data points.

\section{Approach}
\label{sec:approach} 
Now we explain how we compare three steering policies for an atomic clock.
First we explain our choices of steering policies.
Then, we discuss how the numerical simulation works.
Then we elaborate on how we evaluate performance of each policy.
Finally, we specify and justify our choice of noise parameters, weight matrices, and control gains.

We consider the three clock-steering policies
LQG, BB and SMC
discussed in~\S\ref{subsec:cs}.
These three policies are chosen because LQG and BB are the usual standards,
and SMC is what we are introducing for comparison.
The expressions for LQG, BB, and SMC feedback are in Eqs.~(\ref{eq:LQGU}), (\ref{eq:BBu}) and~(\ref{eq:SMCu}), respectively.

Now we explain how the Wiener increments~$\text{d}\bm{W}_k$~(\ref{eq:sdd}) are generated.
We pseudorandomly generate, using the PCG64 version of O'Neill's permuted congruential generator (PCG)~\cite{One14}, two independent sequences of Gaussian Wiener increments~$\text{d}\bm{W}_k$. One sequence is used for the steered clocks and one is used for the reference clock.
The same Wiener increments are reused across SMC, LQG, BB, and the free-running clock so that differences between steering policies are not caused by different noises.
Each component of each sequence is sampled independently from a normal distribution with a mean of zero and variance~$\tau_0$.
The Wiener increments are
\begin{equation}
\label{eq:W_k}
    \text{d}\bm{W}_k := 
    \begin{pmatrix}
        \text{d}W_{1,\,k}\\
        \text{d}W_{2,\,k}
    \end{pmatrix},\;
    \text{d}W_{i,\,k} \sim \mathcal{N}(0,\tau_0),\; i \in \{1,2\}.
\end{equation}
Now that we know how to construct the Wiener increments, we move on to how we iterate the discrete dynamics~(\ref{eq:sdd}) for each steering policy.

For each steering policy, the time series is obtained by solving Eq.~(\ref{eq:sdd}) numerically.
For LQG, we first solve the DARE~(\ref{eq:Riccati}) once to obtain~$\bm{K}^{\text{LQG}}$~(\ref{eq:LQGgain}), then apply~$U_k$~(\ref{eq:LQGU}) through the input vector~$\bm{C}^{X_2}$~(\ref{eq:C_LQG}).
For BB and SMC, we calculate their respective sliding surfaces~(\ref{eq:BBss}, \ref{eq:SMCs}) to compute their respective
$U_k$~(\ref{eq:BBu}, \ref{eq:SMCu}).
Unlike LQG, we apply the feedback law~$U_k$ through the input vector~$\bm{C}^{\dot{X}_2}$~(\ref{eq:C_BBSMC}).
Finally, we iterate the discrete dynamics~(\ref{eq:sdd}) for a finite time~$N\tau_0$, where~$N$ is the total number of discrete-time updates and~$\tau_0$ is chosen to be one day, in agreement with FGTB10.
The numerical solution is thus, for some steering policy $\wp$~(\ref{eq:policies}), 
the entire time series
\begin{equation}
\label{eq:time-ser}
\bm{\mathcal{X}}^s := \{\bm{X}_k^s\}_{k=0}^{N}
\end{equation}
for various seed values
\begin{equation}
\label{eq:seed}
    s \in \{0,1,2 \dots, 99\}.
\end{equation}
Finally,
we use these times series with different seeds~(\ref{eq:time-ser})
to evaluate the performance of the steering policies.

Clock performance is evaluated using two key quantities: accuracy~(\ref{eq:sigmax1}) and stability~(\ref{eq:sigmaytau}).
To evaluate accuracy, we consider four time periods that span short-term, mid-term, and long-term behavior 
\begin{equation}
\label{eq:acc-horizons}
N\tau_0 \in \{1~\text{week},\,1~\text{month},\,1~\text{year},\,10~\text{years}\}.
\end{equation}
For each time period, we calculate~$\bm{\mathcal{X}}^s$ for each of the 100 independent seeds~(\ref{eq:seed}).
Then, we calculate $\sigma_{X_1}$~(\ref{eq:sigmax1}) for each seed and report the mean of~$\sigma_{X_1}$ and its standard deviation across the seeds.
To evaluate stability, we use 
\begin{equation}
\label{eq:stability-horizon}
N\tau_0 = 10^6\,\text{days}
\end{equation}
and seed
\begin{equation}
\label{eq:seedstability}
s = 0.
\end{equation}
Then, we calculate the oADEV~(\ref{eq:sigmaytau}).
A long record for oADEV is needed so that every averaging time
\begin{equation}
\label{eq:ave-time}
\tau = n\tau_0, \quad n \in \mathbb{Z}^+
\end{equation}
utilizes a sufficient number of samples, keeping the overlapping Allan-deviation trace visually clean, i.e., suppressing the significance of oscillatory components.

Now we specify the noise parameters. To ensure that our oADEV and time-offset results are directly comparable to the established reference data, we use the same values
\begin{align}
\label{eq:dc}
    \begin{split}
    \sigma_1 =& 1.02\times10^{-11}\; \sqrt{\text{s}},\\
    \sigma_2 =& 1.97\times10^{-17}\; 1/\sqrt{\text{s}},
    \end{split}
\end{align}
as FGTB10. The reference clock's dynamics is described by Eq.~(\ref{eq:dd}), with
\begin{equation}
\label{eq:sigma_ref}
(\sigma_{1,\,\text{ref}},\,\sigma_{2,\,\text{ref}})
= \alpha
(\sigma_{1},\,\sigma_{2}),
\end{equation}
where
\begin{equation}
\label{eq:alpha}
\alpha\in\{0.03, 0.10, 0.30\}.
\end{equation}
We repeat the accuracy analysis for different values of~$\alpha$ to test the sensitivity of our conclusions to the assumed reference-clock quality.
The~$\alpha = 0.10$ case is consistent with the benchmarking scenario in FGTB10.

The weight matrices and control gains are chosen as follows.
The DARE~(\ref{eq:Riccati}) is solved with the weight matrices
\begin{equation}
\label{eq:WQ}
W_Q := \text{diag}(\tau_0^{-2}, 1),\,
W_R = 10^2.
\end{equation}
These weights reproduce ``Case 2” in FGTB10, which is the variant that gives the best compromise between accuracy and short-term stability. The gain~$K_{\text{BB}}$ is taken to match the value used in FGTB10
\begin{equation}
\label{eq:bbgain}
    K_{\text{BB}} = 1.0 \times10^{-19} \;\text{s}^{-1},
\end{equation}
whereas~$K_{\text{SMC}}$ is tuned by trial and error to minimize both the time-offset standard deviation and the short-term instability
\begin{equation}
\label{eq:SMCgain}
    K_{\text{SMC}} = 1.1\times10^{-19} \;\text{s}^{-1}.
\end{equation}
Finally, similar to~$K_{\text{SMC}}$, via trial and error we set
\begin{equation}
\lambda = 6 \times 10^{-6} \;\text{s}^{-1}
\end{equation}
as the decay rate for SMC policy.

\section{Results}
\label{sec:res}
In this section, we present our results. 
We begin by comparing the time-offset
trajectories of the steered clocks and a FRC over the first 250 days
for a particular seed number $s$~(\ref{eq:seedstability}).
Second, we quantify accuracy of the steered clocks and the FRC based on the mean and standard deviation of variance~(\ref{eq:sigmax1}) across multiple seeds and over multiple time periods.
Third, we discuss the robustness of our results to the assumed reference-clock quality.
Finally, we assess clock stability by
analyzing the oADEV~(\ref{eq:sigmaytau}) across a range of averaging times.

\begin{figure}
  \centering
  \includegraphics[width=\linewidth]{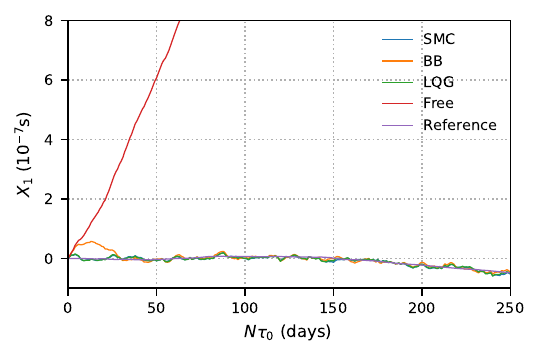}
  \caption{%
Time-offset trajectories \(X_1(t)\) in
$10^{-7} \times$ vs time of evolution in days
over the first 250 days for five cases, coloured according to the legend: SMC, BB, LQG, FRC, and the reference clock for~$\alpha = 0.10$}.
\label{fig:timeoffset}
\end{figure}

Now we compare the clocks' time-offset trajectories. 
Figure~\ref{fig:timeoffset} 
shows the first 250 days of the time-offset data for the three steered clocks and the FRC for a particular seed~(\ref{eq:seedstability}).
We predict that all steered clocks' time to remain within $\pm 0.1~\upmu$s of the reference for the entire 250 days, whereas we predict that the FRC's time will drift monotonically. 
In the first few tens of days, the BB trajectory departs more visibly from the reference, whereas SMC and LQG follow the reference clock from the start. After that transient, all three steered clocks follow the reference clock.

We next quantify the time-domain accuracy of each steering policy.
Table~\ref{tab:acc} 
\begin{table}
\centering
\begin{tabular}{c
                r@{$\,\pm\,$}l
                r@{$\,\pm\,$}l
                r@{$\,\pm\,$}l
                r@{$\,\pm\,$}l}
\toprule
Policy
& \multicolumn{2}{c}{1 week}
& \multicolumn{2}{c}{1 month}
& \multicolumn{2}{c}{1 year}
& \multicolumn{2}{c}{1 decade} \\
\midrule
SMC
& 4.2  & 1.4
& 5.6  & 1.8
& 5.91 & 0.69
& 5.95 & 0.26 \\
LQG
& 4.4  & 1.5
& 5.9  & 1.7
& 5.80 & 0.40
& 5.76 & 0.15 \\
BB
& 17.3 & 3.0
& 24.7 & 4.0
& 16.3 & 3.9
& 9.87 & 0.77 \\
FRC
& 23.2 & 3.1
& 91   & 15
& 1130 & 550
& 21000 & 14000 \\
\bottomrule
\end{tabular}
\caption{Accuracy performance for each time period and each steering policy for~$\alpha = 0.10$. The numbers represent the mean $\pm$ the standard deviation of $\sigma_{X_1}$~(ns) over 100 different seeds for three steered clocks and the FRC.}
\label{tab:acc}
\end{table}
reports the mean of~$\sigma_{X_1}$ plus or minus its standard deviation over various seeds~(\ref{eq:seed}) for each policy and time period for $\alpha = 0.10$~(\ref{eq:alpha}).
This table shows that SMC has the lowest mean~$\sigma_{X_1}$ at one week and one month, whereas LQG has the lowest mean~$\sigma_{X_1}$ at one year and one decade. 
The difference between SMC and LQG is small compared to the improvement of either method over BB.
Both SMC and LQG outperform BB for every time period considered, and the FRC performs worst by a wide margin.

We now discuss the robustness of our results to the assumed reference-clock quality. For all tested values of $\alpha$~(\ref{eq:alpha}), we find that SMC remains slightly more accurate than LQG at one week and one month, whereas LQG is slightly more accurate at one year and one decade. For every~$\alpha$, both SMC and LQG remain substantially more accurate than BB. Thus, the detailed ordering of SMC and LQG depends on the considered time period, but the conclusion that SMC is competitive with LQG and substantially better than BB is robust.

\begin{figure}
  \centering
  \includegraphics[width=\linewidth]{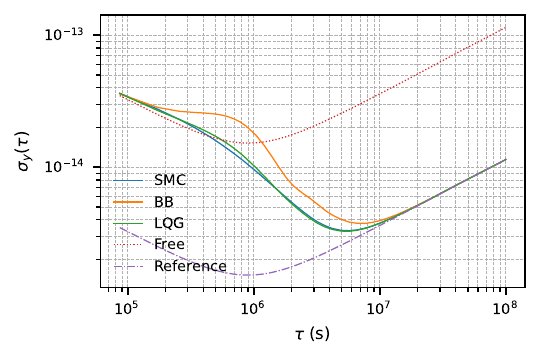}
  \caption{Overlapping Allan deviation of the time-offsets for five cases, coloured according to the legend: SMC, BB, LQG, free-running, and the reference clock for~$\alpha = 0.10$.}
  \label{fig:alldev}
\end{figure}

We then examine how the steering strategies affect clock stability.
Figure~\ref{fig:alldev} 
shows the overlapping Allan deviation, $\sigma_y(\tau)$, for averaging times given in Eq.~(\ref{eq:Allandeviation}). 
The LQG and SMC curves lie nearly on top of one another across the entire domain. 
Both LQG and SMC show modest short-term instability.
Beyond this short-term behaviour, both curves converge with the reference at the random-walk floor.
The BB curve departs from LQG and SMC at 
\begin{equation}
\label{eq:tauapprox}
\tau \approx 2 \times 10^5\,\text{s}
\end{equation}
peaking around $6 \times 10^{5}\,\text{s}$ before converging back toward the common RWFN slope beyond~$10^7\,\text{s}$. 
For completeness, the FRC’s~$\sigma_y(\tau)$ is also shown: it intersects the LQG and SMC steered curves at $\tau \approx 4 \times 10^5\,\text{s}$ and the BB curve at $\tau \approx 10^6\,\text{s}$, then continues to rise with a slope of~$+\nicefrac1{2}$, which is the characteristic of an uncompensated random-walk process~$\cite{RH08}$.
We also checked the oADEV behavior for all the~$\alpha$ values; although the detailed curves change with~$\alpha$, the qualitative conclusion is unchanged: SMC remains close to LQG in stability and avoids the short-term instability seen in BB.

\section{Discussion}
\label{sec:dis}
Now we discuss our results. First, we discuss the time-offset trajectories to give an intuitive picture of how each steering policy keeps the clock near the reference. 
Second, we discuss the accuracy comparison to summarize how the steering policies rank when averaged over many noise realizations. 
Finally, we discuss the stability comparison to describe how the steering policies influence the clock’s stability across averaging times.

We first discuss the time-offset trajectories.
We analyse time-offset trajectories to elucidate the connection between the steering policy and the clock accuracy.
The steering loop for a given policy keeps a clock from drifting away from its reference clock.
Our trajectories show that SMC and LQG behave similarly, whereas BB exhibits visibly abrupt corrections initially because it acts like a relay, meaning the control switches between fixed correction values, rather than acting as a continuously adjusting controller. However, BB approaches the SMC and LQG at longer times.

We next discuss which steering policy is most accurate, i.e., closest to the reference clock, based on our simulations.
By simulating the noisy evolution of a clock steered by our three policies,
we determine that SMC and LQG provide the best average accuracy, with their detailed ordering depending on the time period considered.
Our simulations show that SMC is slightly more accurate than LQG at one week and one month, whereas LQG is slightly more accurate at one year and one decade.
The BB is consistently less accurate,
and the unsteered clock is worst.
Importantly, our simulations over many noise realizations and reference-clock noise levels show that our ranking of steering policies is not a numerical artifact but rather a genuine system effect.
Practically speaking, this means that SMC achieves excellent timekeeping without requiring the detailed model-based machinery associated with LQG, and SMC does so robustly despite the choice of random seed and reference-clock quality in our simulation.

Finally, we discuss the stability comparison.
The stability comparison addresses how the steering policies affect the clocks' stability. The main message is that SMC preserves essentially the same stability behaviour as LQG across the full averaging-time range studied, whereas BB displays a distinct short-term instability associated with its intermittent, relay-like corrections. 
At long averaging times, all steered clocks ultimately follow the stability set by the reference clock, meaning that, at long averaging times, the stability of the reference is the limiting factor rather than the details of the steering policy.
In practical terms, SMC provides a stability performance comparable to LQG, while avoiding the characteristic short-term instability seen in BB; this conclusion holds for all tested reference-clock qualities.
\section{Conclusion}
\label{sec:con}
Our study sets out to determine whether a continuously active, first-order sliding-mode controller~(SMC) could serve as a practical long-term steering policy for atomic clocks. 
For that purpose,
we analyzed how SMC performance compares with two established alternatives: optimally tuned LQG and BB.
Using a common two-state clock model driven by WFN and RWFN, we evaluated time-domain accuracy via Monte-Carlo simulations over many noise realizations and assessed stability using overlapping Allan deviation across the full averaging time span of interest.
By these methods, we have shown that SMC,
governed by an appropriate control policy,
could lead to the SMC having comparable accuracy with LQG without causing a short-term instability seen by BB.

Looking ahead, the main practical question is whether the simulated competitiveness of SMC can be converted into an operational advantage in real clock-steering systems. 
Our results do not imply that existing LQG or BB steering policies should immediately be replaced; rather, our results suggest that SMC is a promising candidate for experimental testing because its implementation cost is potentially low. 
In our framework, SMC uses the same clock-state information as the other steering policies and requires only a simple sign-based nonlinear correction with only two tunable parameters. Therefore, if an existing steering policy already estimates the time and frequency offsets, testing SMC could be a software-level modification rather than a hardware-level redesign. 
Future work should therefore test SMC using real clock data while accounting for measurement delays, limits on the applied steering commands $U_k$~(\ref{eq:LQGU}, \ref{eq:BBu}, \ref{eq:SMCu}), imperfect state estimation, and broader noise models to determine whether the cost-benefit trade-off remains favourable outside simulation.
\acknowledgments
We acknowledge support from the Natural Sciences and Engineering Research Council of Canada (NSERC) and the Government of Alberta.
We thank Ingo Nosske for helpful comments.
We acknowledge the traditional territories of the people of Treaty 6 and Treaty 7 in Alberta.
\bibliography{Manuscript}
\end{document}